# ESTAN: Enhanced Small Tumor-Aware Network for Breast Ultrasound Image Segmentation

Bryar Shareef, Alex Vakanski, *Member, IEEE,* Min Xian, *Member, IEEE, Phoebe E. Freer*

***Abstract*—Breast tumor segmentation is a critical task in computer-aided diagnosis (CAD) systems for breast cancer detection because accurate tumor size, shape and location are important for further tumor quantification and classification. However, segmenting small tumors in ultrasound images is challenging, due to the speckle noise, varying tumor shapes and sizes among patients, and existence of tumor-like image regions. Recently, deep learning-based approaches have achieved great success for biomedical image analysis, but current state-of-the-art approaches achieve poor performance for segmenting small breast tumors. In this paper, we propose a novel deep neural network architecture, namely Enhanced Small Tumor-Aware Network (ESTAN), to accurately and robustly segment breast tumor. ESTAN introduces two encoders to extract and fuse image context information at different scales and utilizes row-column-wise kernels in the encoder to adapt to the breast anatomy. We validate the proposed approach and compare to nine state-of-the-art approaches on three public breast ultrasound datasets using seven quantitative metrics. The results demonstrate that the proposed approach achieves the best overall performance and outperforms all other approaches on small tumor segmentation.**

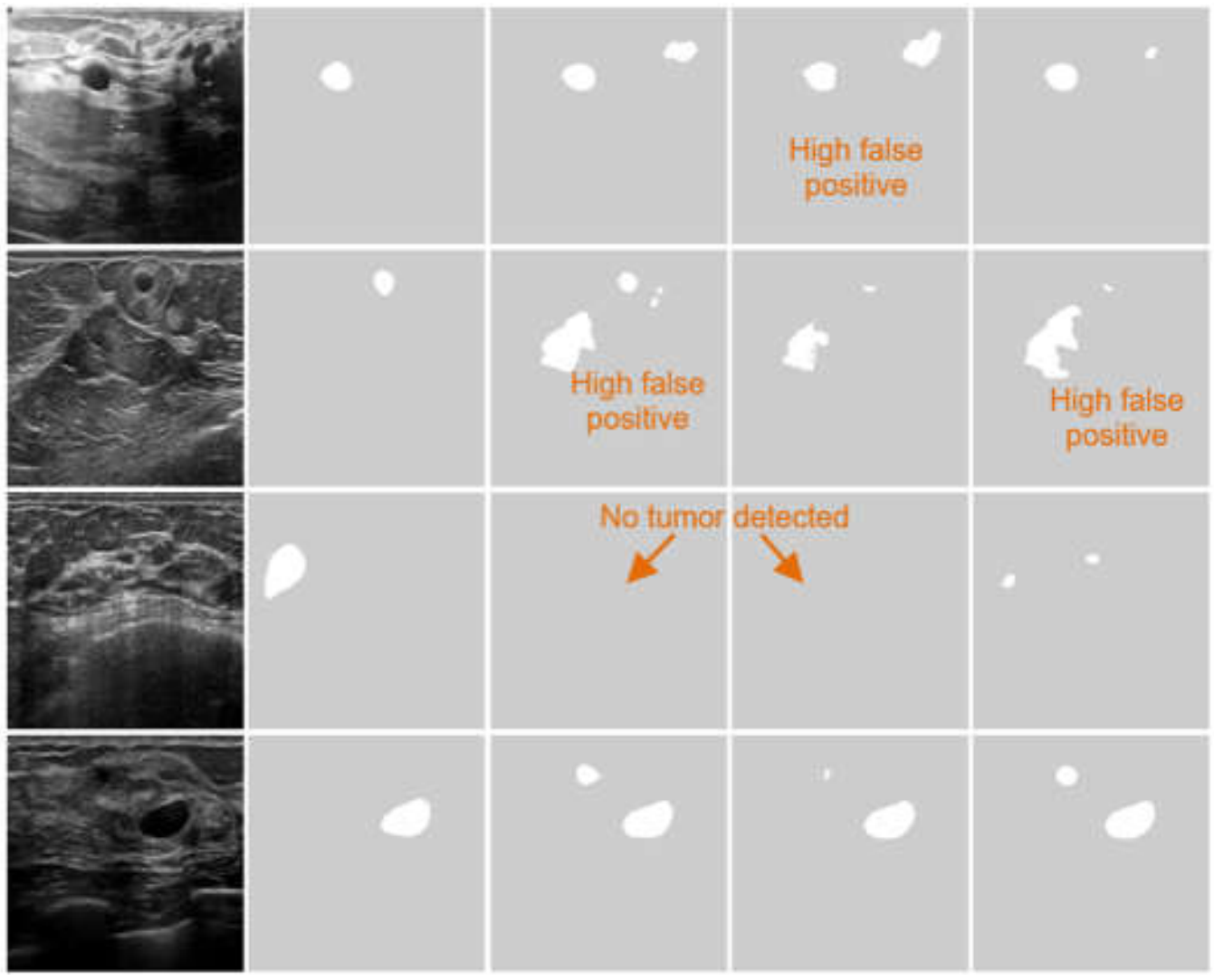

(a) BUS Images (b) GT (c) DenseU-Net (d) CE-Net (e) RDAU-Net

Fig. 1. Performance of state-of-the-art approaches for segmenting breast tumors with different sizes. GT: Ground truth.

## I. INTRODUCTION

BREAST ultrasound (BUS) imaging has become an effective screening method due to its painless, noninvasive, nonradioactive and cost-effective nature. BUS image segmentation aims to extract tumor region(s) from normal breast tissues in images. It is an essential step in BUS computer-aided diagnosis (CAD) systems. However, because of the speckle noise, poor image quality and variable tumor shapes and sizes, accurate BUS image segmentation is challenging.

According to the National Cancer Institute in the United States, the relative survival is 99% if the breast cancer is detected and treated at the early stages, and only 27% if the cancer has spread to other organs of the body[1]. Early detection of breast tumor is the key to reducing the mortality rate. However, at the early stages, most tumor are small and occupy a relatively small region in BUS images. It is challenging to distinguish them from normal breast tissues. Therefore, accurate detection of small tumors is critical for breast cancer early detection, and can improve clinical decision, treatment planning, and recovery.

The approaches of BUS image segmentation can be classified into traditional approaches and deep learning-based approaches. Numerous traditional approaches have been used to BUS image segmentation, such as thresholding [2][3][4][5][6][7], region growing [8][9], and watershed [10]. Despite their simplicity, these methods require knowledge and expertise in extracting features, and they are not robust due to poor scalability and high sensitivity to noise. Refer to [11] for a comprehensive review of BUS image segmentation.

Recently, several deep learning approaches [12]-[21] have been developed for BUS image segmentation; TABLE I lists the most recent deep learning approaches for BUS image segmentation. Huang *et al.* [12] proposed a fuzzy fully convolutional network to perform BUS image segmentation. Contrast enhancement and wavelet features were applied as a preprocessing approach to augment the training data. The augmented training image set and features from convolutional layers were transformed to a fuzzy domain by a fuzzy membership function. The context information and the human breast structure are integrated to the Conditional Random Fields (CRFs) to enhance the segmentation results. Yap *et al.* [13]

This work was supported by the Institute for Modeling Collaboration (IMCI) at the University of Idaho through NIH Award #P20GM104420. (Corresponding author: Min Xian)

B. Shareef, A. Vakanski, and M. Xian are with the Department of computer science, University of Idaho at Idaho Falls, Idaho Falls, ID 83401 USA (e-mails: shar0416@vandals.uidaho.edu, vakanski@uidaho.edu, mxian@uidaho.edu).

P. Freer is with the Department of Radiology and Imaging Sciences, University of Utah School of Medicine, Salt Lake City, UT 84132, USA (e-mail: phoebe.freer@hsc.utah.edu)

TABLE I
DEEP LEARNING-BASED BUS SEGMENTATION APPROACHES

| Article | Year | Method | Dataset Size | Evaluation Metrics* |
|---|---|---|---|---|
| Huang *et al.* [12] | 2018 | FCN + Wavelet features + CRFs | 325 | TPR, FPR, JI |
| Yap *et al.* [13] | 2018 | Patch-based LeNet, U-Net, and FCN-AlexNet | 469 | TPR, FPR, F1 |
| Zhuang *et al.* [14] | 2019 | U-Net+ Attention gate | 1062 | TPR, Sp, F1, Pr, JI, Acc, DSC, AUC |
| Hu *et al.* [15] | 2019 | Dilated FCN + Active contour model | 570 | DSC, MAD, and HD |
| Vakanski *et al.* [16] | 2020 | U-Net + Attention blocks | 510 | TPR, FPR, DSC, JI, Pr, AUC-ROC |
| Lee *et al.* [19] | 2020 | U-Net + Channel attention module | 163 | FPR, F1, JI, AUC, Pr, Sp, TPR |
| Moon *et al.* [18] | 2020 | Ensemble CNNs | 246 | TPR, FPR |
| Byra *et al.* [17] | 2020 | U-Net + Attention gate + Entropy maps | 269 | DSC, JI |
| Shareef *et al.* [24] | 2020 | U-Net + Two encoders | 725 | TPR, FPR, JI, DSC, AER, MAE, AHE |

*TPR: true positive rate, FPR: false positive rate, JI: Jaccard indices, IoU: intersection over union, Acc: Accuracy, Pr: precision, Sp: specificity, MCC: mattews correlation coefficient, AUC: area under curve, AER: area error rate, MAE: mean area error, AHE: average Hausdorff error, DSC: dice similarity coefficient, CRFs: conditional random fields, and FCN: fully convolutional network.

evaluated the performance of three different deep learning approaches for segmenting BUS images: a patch-based LeNet, a U-Net, and a transfer learning with a pretrained AlexNet. These three methods achieved remarkable overall performance in segmenting BUS images on two different datasets. Zhuang *el at.* [14] proposed an RDAU-Net model, based on U-Net architecture, to perform the tumor segmentation task on BUS images, where dilated residual blocks and attention gates were used to replace the basic blocks and original skip connections in U-Net, respectively. Similarly, Hu *et al.* [15] proposed a method that combined the dilated fully convolution network with a phase-based active contour model. Moreover, to exclude tumor-like regions, the method in [16] integrated radiologists' visual attention for BUS segmentation. Byra *et al.* [17] proposed a deep learning segmentation approach based on entropy parametric maps. The attention gate block is employed to improve the performance of the segmentation task. Furthermore, Moon *et al.* [18] proposed an ensemble CNN architecture for CAD system to diagnose BUS images. The ensemble approach comprises multi-models where each is trained on original images, segmented image tumors, tumor masks, and fused images. The fused images were prepared by combining an original image, segmented tumor, and the tumor shape information (TSI). Lee *et al.* [19] proposed channel attention module with multi-scale grid average pooling for segmenting BUS images. The approach utilizes both local and global information to improve the segmentation performance. These methods achieved good overall performance. However, as shown in Fig. 1, they failed to achieve good performance for segmenting small tumors. First, these methods are designed to improve overall performance using general-purpose square kernels which are developed to learn features from natural images. Second, all currently available BUS datasets are small, and most deep learning-based approaches require a large and high-quality training set.

Small object detection and/or segmentation is challenging in computer vision. It forms the foundation of many image related tasks, such as remote sensing, scene understanding, object tracking, instance and panoptic segmentation, aerospace detection, and image captioning. Chen *et al.* [20] proposed an augmented technique for the R-CNN algorithm with a context model and small region proposal generator; which was the first benchmark dataset for small object detection. Krishna *et al.* [21] designed a Faster R-CNN with a modified upsampling technique to improve the performance of small object detection. Guan *et al.* [22] proposed a semantic context aware network (SCAN), which integrates location fusion module and context fusion module to detect semantic and contextual features. The DenseU-Net architecture was proposed by Dong [23], which performs semantic segmentation of small objects in urban remote sensing images. It uses residual connections and a weighted focal loss function with median frequency balancing to improve the performance of small object detection.

To the best of our knowledge, STAN [24] was the first deep learning architecture to improve small tumor segmentation. Three skip connections and two encoders were employed to extract multi-scale contextual information from different layers of the contracting part. STAN outperformed other deep learning approaches for segmenting small tumors in BUS images. However, its average false positive rate on small tumors is much larger than that of large tumors. In this paper, we extend STAN and propose a new architecture, namely Enhanced Small Tumor-Aware Network, to achieve robust segmentation for tumors with different sizes. The new architecture has two encoder branches. The basic encoder has five blocks and learns features at different scales. The ESTAN encoder applies row-column-wise kernels to adapt to the breast anatomy during the feature learning. In the decoder, each block has three skip connections that fuse rich contextual features from the two encoders. The contextual features are robust to different tumor sizes and help distinguish tumor regions from normal regions.

The rest of the paper is organized as bellow: Section II presents the proposed architecture and implementation details; Section III demonstrates experimental results; and Section IV

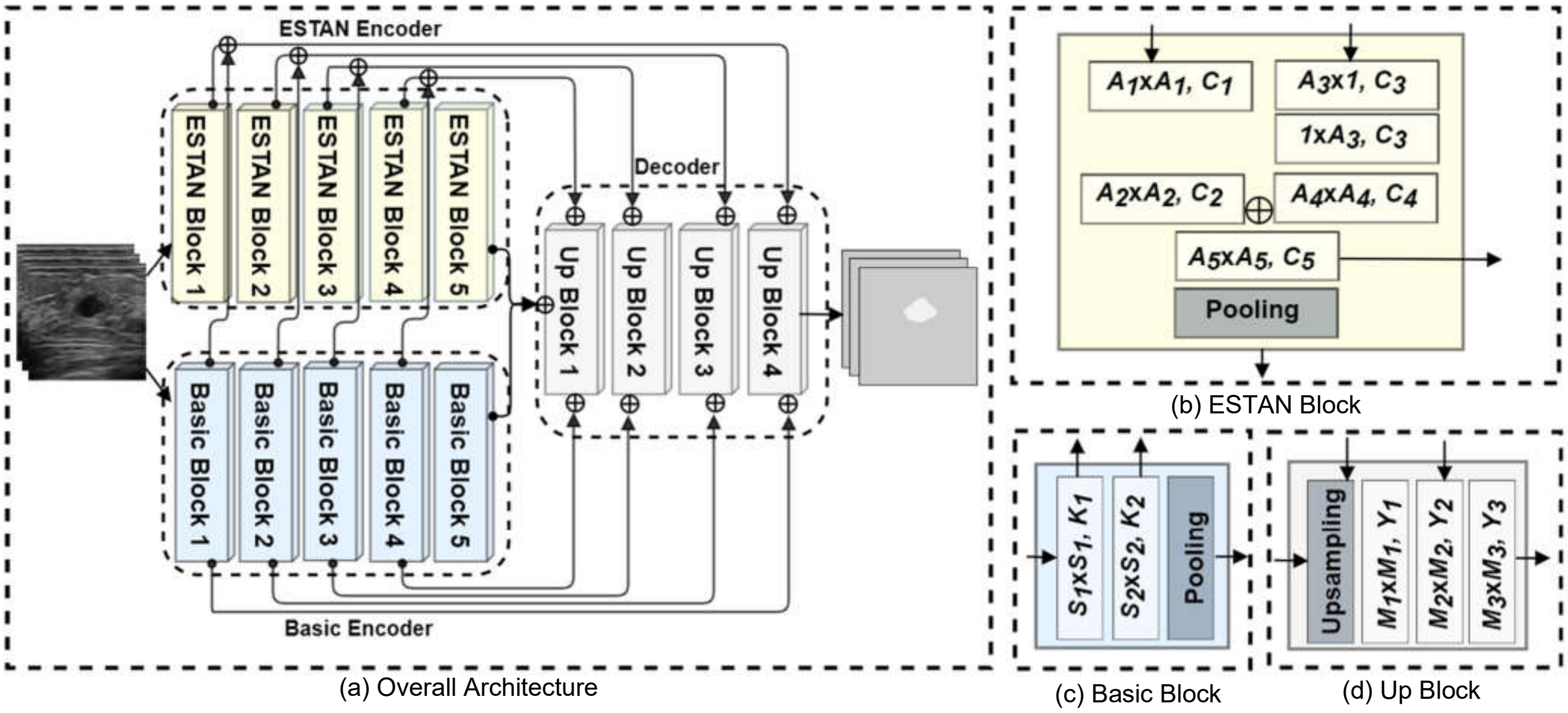

Fig. 2. ESTAN architecture. ⊕ is the concatenation operator, $\boldsymbol{A_i}$, $\boldsymbol{S_i}$, $\boldsymbol{M_i}$, denote kernel sizes, and $\boldsymbol{C_i}$, $\boldsymbol{K_i}$, $\boldsymbol{Y_i}$ define number of kernels.

provides the conclusion and discusses the future work.

## II. Enhanced Small Tumor-Aware Network

In this section, we introduce the proposed Enhanced Small Tumor-Aware Network (ESTAN) for solving the issue of small tumor segmentation in BUS images. ESTAN builds upon two observations: 1) BUS images contain tumors of a broad range of sizes, and current state-of-the-art approaches have poor performance on segmenting small tumors; and 2) the current deep learning-based approaches used square-shape kernels and have difficulty utilizing context information of BUS images, e.g., breast tissue anatomy. To alleviate these challenges, we propose the ESTAN to extract and fuse image context information at different scales. ESTAN constructs feature maps using both square and large row-column-wise kernels. These feature maps transmit multi-scale context information and preserve fine-grained tumor location information. Therefore, the new design enables ESTAN to accurately segment breast tumors of different sizes, and it is especially efficient with small size tumors. ESTAN consists of two encoders and one decoder with three skip connections. The overall architecture of the proposed approach is shown in Fig. 2.

### *A. Basic Encoder*

The basic encoder down-samples the input feature maps to extract low-level spatial and contextual information. Both convolution and pooling operations with strides greater than 1 are employed for downsampling the feature maps in the encoder blocks. The basic encoder comprises of five blocks, where each block contains two convolutional layers and a max pooling layer; except the fifth block, which has no pooling layer. The basic blocks in the encoder are different from the original U-Net encoder blocks, since the new architecture uses two skip connections to copy feature maps from the encoder blocks to the corresponding upsampling layers in the decoder module. Fig. 2(c) illustrates the architecture of the basic encoder. Let denote the input images as $X \in \mathbb{R}^{h\times w\times c}$, where $h$, $w$ and $c$ are the height, width, and number of channels, respectively. Let $f$ be the convolution function for square kernels, $K_i$ be the number of kernels and $S_i$ be kernel size in the $i$th convolution layer, respectively. The output of the $j$th block of the basic encoder is defined by

$$B_j = \phi\left(f_{S_2,K_2}\left(f_{S_1,K_1}(X)\right)\right) \quad (1)$$

where $B_j$ is the output of a given block, and $\phi$ is the pooling operation in the $j$th block. Additionally, $K_1, K_2, K_3, K_4$, and $K_5$ have values 32, 64, 128, 256, and 512, respectively.

### *B. ESTAN Encoder*

The receptive field in CNNs has an important role in building effective feature maps. It defines the input image region that produces the output feature, and image regions outside the receptive field of a feature will not contribute to the computation of the feature. To ensure the coverage of all relevant image regions and achieve enhanced performance, many dense prediction tasks used large receptive fields [25][26]. There are several techniques for increasing the size of the receptive field such as stacking more layers, sub-sampling, and dilated convolutions [27]. However, in BUS image, large receptive field can result in poor performance for small tumors segmentation [24]. The goal ESTAN encoder is to produce effectively feature maps and avoid large receptive field.

STAN [24] proposed a two-encoder architecture and only applied small kernels, e.g., $1\times1, 3\times3,$ and $5\times5$. The small kernels can avoid large receptive field. The two encoders fused contextual information at different scales by producing features using different sizes of receptive fields. This design improved the overall performance for small breast tumor segmentation. However, STAN produced high false positive for BUS images with some small tumors.

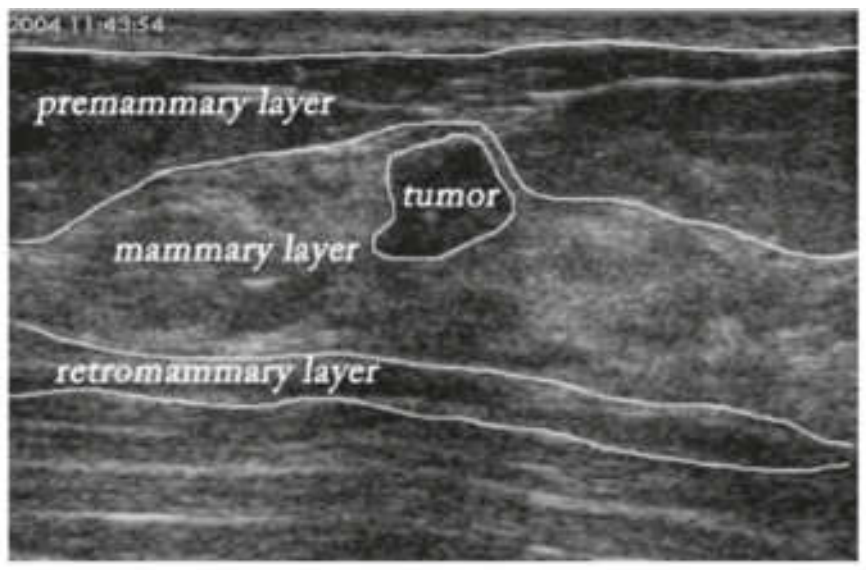

Fig. 3. Major breast layers.

To overcome this problem, we redesign the encoder by applying row-column-wise kernels. The small square kernels in STAN constructed feature maps using only using square image regions. The motivation of the design is that BUS images are composed of vertically stacked breast layers (Fig. 3). Applying row-column-wise kernels in CNNs can avoid calculating features using images regions from multiple anatomical layers and produce more accurate and meaningful feature maps.

ESTAN encoder comprises five blocks, named ESTAN blocks, which are parallel with the basic encoder blocks. Each block has four square kernels and two row-column-wise kernels in two parallel branches. Such kernels can efficiently extract contextual and fine-grained details of small tumors in the BUS images. Furthermore, ESTAN blocks add one extra non-linearity to each encoder blocks. Fig. 2(b) illustrates the design of ESTAN block. Let $C_i$ be the number of kernels, and $A_i$ be the kernel size. The output of $j$th ESTAN block is defined by

$$E_j = \phi\left(f_{A_5,C_5}\left(f_{A_2,C_2}\left(f_{A_1,C_1}(X)\right) + f_{A_4,C_4}\left(h_{1,A_3,C_3}\left(h_{A_3,1,C_3}(X)\right)\right)\right)\right) \quad (2)$$

where $E_j$ is the output of the $j$th ESTAN block, and $\phi$ is the pooling operation, $h$ is the row-column-wise convolution function with the size of $A_3 \times 1$ and $1 \times A_3$, respectively. The size of $A_3$ in $E_1, E_2, E_3, E_4$, and $E_5$ blocks are 15, 13, 11, 9, and 7, respectively. The size of $A_5$ in $E_2$ and $E_5$ is 5, and in the rest is 1. Furthermore, Block 5 has no pooling operation for both encoders. Moreover, $C_1, C_2, C_3, C_4$, and $C_5$ have values 32, 64, 128, 256, and 512, respectively.

In addition, STAN has 22 million parameters while ESTAN uses 30 million, because ESTAN uses more convolution layers in both encoder and decoder. The training time for both STAN and ESTAN is fast, and it depends on the dataset size, batch size, and the hardware specification of the machine.

### C. Decoder and Skip Connections

The decoder module comprises four upsampling blocks, where each has one upsampling followed by three convolution layers. Unlike the U-Net architecture, where the decoder has two convolution layers, the ESTAN adds an additional kernel after the first convolution kernel to control the post concatenation channels. Let $f$ be the convolution function, $Y_i$ be the number of kernels, and $M_i$ be the kernel size. The output of the $j$th block of the decoder is defined by:

$$U_j = f_{M_3,Y_3}\left(f_{M_2,Y_2}\left(f_{M_1,Y_1}(\Psi)\right)\right) \quad (3)$$

where $\Psi$ is the upsampling layer. $M_1$ and $M_3$ in all blocks are 3 and $M_2$ in block 1,2, and 3 is 1, and $M_2$ in block 4 is 5.

To overcome the singularity issues during the training, we have introduced three skipping connections to copy feature maps at different scales from both encoders to the decoder. The possible singularities that occur are overlap, elimination and linear dependence singularities. The first two skip connections come from combining the result of $f_{S_1,K_1}$ in the basic encoder and the result of $f_{A_5,C_5}$ in the ESTAN encoder concatenates to the upsampling layer. The second skip connection that comes from the result of $f_{S_2,K_2}$ combines to the $f_{M_2,Y_2}$ in the decoder part. In addition, $Y_1, Y_2, Y_3$, and $Y_4$ are 256, 128, 64, and 32, respectively. Fig.2(d) illustrates the decoder module.

### D. Implementation and Training

In this work, we use three public datasets [28][29][13] to train and test all the approaches. The input images and their ground truths are resized to $256 \times 256$ pixels. We applied image width and height shift augmentation techniques to the training set of Dataset B, which has only 163 BUS images. During the training, the batch size is set to 4 and the maximum number of epochs is set to 50. To train the model, we applied adaptive moment estimation (Adam) [30], and the initial learning rate is set to 0.0001. In most BUS images, the number of the tumor pixels is much smaller than that of background pixels, which might cause the overclassification the background pixels. To alleviate this issue, we employed the Dice loss [31] to measure the relative overlap between the ground truth and the predicted labels. The dice loss function $D_l$ is defined by

$$D_l = 1 - \frac{1 + 2\sum_i^N p_i g_i}{1 + \sum_i^N p_i^2 + \sum_i^N g_i^2} \quad (4)$$

where $P = \{p_i \in [0,1]\}_{i=1}^N$ and $G = \{g_i\}_{i=1}^N$ are the output of the final pixel-wise sigmoid layer and the ground truth, respectively.

## III. Experimental Results

### A. Datasets, Evaluation Metrics and Setup

We use three public BUS datasets: BUSIS dataset [28], BUSI dataset [29] and Dataset B [13].

The BUSIS dataset contains 562 images collected from three hospitals using GE VIVID 7, LOGIQ E9, Hitachi EUB-6500, Philips iU22, and Siemens ACUSON S2000. The BUSI dataset is from Baheya Hospital for Early Detection & Treatment of Women's Cancer in Egypt using LOGIQ E9 ultrasound system and LOGIQ E9 Agile ultrasound system with the ML6-15-D Matrix linear probe transducers. The BUSI dataset has 780 images, of which there are 133 normal, 487 benign, and 210 malignant images. The Dataset B has only 163 breast ultrasound images, and the UDIAT Diagnostic Centre of the Parc Taul´ı Corporation, Sabadell (Spain) collected the images using Siemens ACUSON Sequoia C512 system with 17L5 linear array transducer.

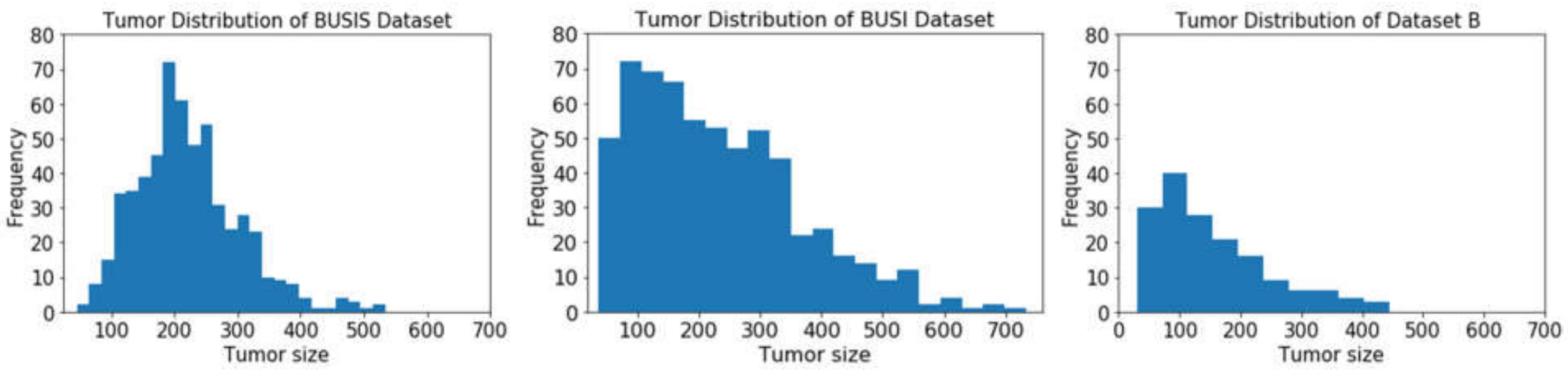

Fig. 4. Histogram of tumor size (number of pixels) distribution per dataset.

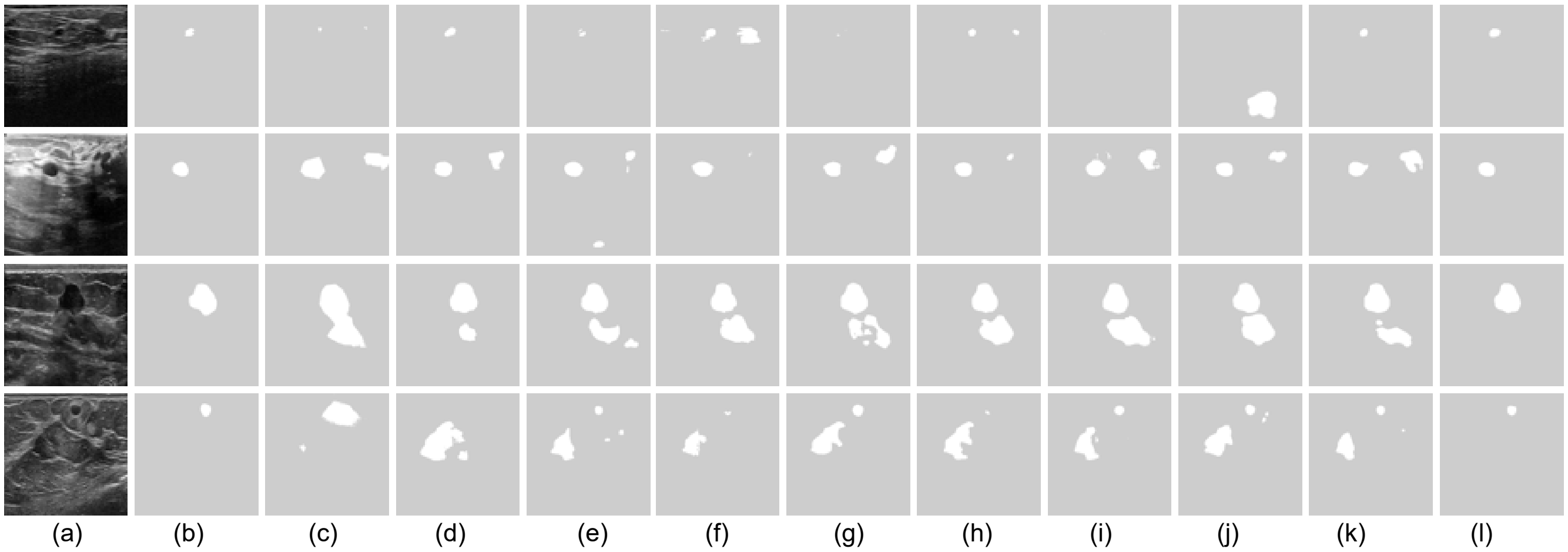

Fig. 5. Tumor segmentation examples. (a) BUS Image, (b) ground truth, (c) AlexNet, (d) SegNet , (e) U-Net, (f) CE-Net, (g) MultiResUNet, (h) RDAU-Net , (i) SCAN, (j) DenseU-Net, (k) STAN, and (l) ESTAN.

The tumor size is an important variable, and Fig. 4 illustrates the histograms of tumor size distributions of the three datasets based on their original BUS image. The physical sizes of most tumors of the three datasets are unavailable; therefore, we define the tumor size as the length (pixels) of the longest axis of a tumor region in original BUS image. The distributions of BUSI and Dataset B show skewed shapes to the right where many tumors are smaller than 150 pixels. The BUSI dataset has more large tumors compared to the other datasets, and the sizes of most tumors are from 150 and 250 pixels. In addition, BUSIS dataset are from five different BUS workstations and the image quality has large variations.

To evaluate the segmentation results, both area and boundary metrics are employed. The metrics are true positive rate (TPR), false positive rate (FPR), Jaccard index (JI), Dice similarity coefficient (DSC), area error rate (AER), Hausdorff error (HE) and mean absolute error (MAE). For detailed information about the seven metrics, refer to [28]. We perform 5-fold cross-validation to evaluate the test performance of all methods, and the input image size is $256 \times 256$ pixels for all the approaches. In this study, we compare the proposed method with nine state-of-the-art approaches: AlexNet [32], SegNet [33], U-Net [34], CE-Net [35], MultiResUNet [36], RDAU-Net [14], SCAN [22], DenseU-Net[37], and STAN [24]. These approaches have different backbone networks and different training strategies. We employed transfer learning technique for the AlexNet, which is pretrained on ImageNet. The SegNet, AlexNet, U-Net, CE-Net, MultiResUNet, RDAU-Net, SCAN, DenseU-Net are trained from scratch.

All approaches are tested using a workstation with a 3.50 GHz Intel(R) Xeon(R) CPU, a 32 GB of ram, and an Nvidia Titan Xp GPU.

## B. Overall Performance

In this section, we compare the proposed approach with AlexNet, SegNet, U-Net, CE-Net, MultiResUNet, RDAU-Net, SCAN, DenseU-Net, and STAN. The results are shown in Fig. 5 and Table II.

Fig. 5 shows the segmentation results of four sample BUS images. In the first row, the tumor in the BUS image is small, and AlexNet, U-Net, MultiResUNet, SCAN and DenseU-Net have poor segmentation performance. In the second and third samples (2nd and 3rd rows), all approaches, except the proposed ESTAN, produce high false positives, which demonstrates that they have difficulty in distinguishing tumor region from tumor-like regions. In Fig. 5(k), STAN can segment small tumors accurately, but still produce false tumor regions. Fig. 5(l) shows that ESTAN segments the four images accurately without any false tumor regions.

TABLE II illustrates the overall quantitative results of all approaches on three datasets. The proposed ESTAN achieves the best overall performance on all three datasets. AlexNet and SegNet obtain high TPRs, but at the cost of high FPRs.

## C. Small Tumor Segmentation

The physical size for all images of the three datasets are not available. Therefore, the length of the longest axis of a tumor region from original BUS image (non-resized) is chosen to be a criterion to select small tumors, and the length threshold is set to 120 pixels. BUSIS, BUSI, and Dataset B contain 49, 151, and 76 small tumors, respectively. Fig. 6 illustrates the false positive rate comparison between the overall and small tumor segmentation. All ten approaches have higher false positive rate for small tumors. The false positive rate of AlexNet has increased dramatically for small tumor segmentation. The ESTAN approach is superior in comparison to all nine approaches and achieves the lowest false positive for both overall and small tumor segmentation. TABLE III shows all-inclusive results of all approaches on the three datasets using seven quantitative metrics. ESTAN outperforms all other nine approaches for small tumor segmentation on the three datasets. AlexNet and SegNet obtain high TPRs, but at the cost of high FPR.

TABLE II
OVERALL PERFORMANCE

| Datasets | Methods | TPR | FPR | JI | DSC | AER | AHE | AME |
|---|---|---|---|---|---|---|---|---|
| BUSIS [28] | AlexNet | **0.95** | 0.34 | 0.74 | 0.84 | 0.39 | 25.1 | 7.1 |
| | SegNet | 0.94 | 0.19 | 0.82 | 0.90 | 0.22 | 21.7 | 4.5 |
| | U-Net | 0.92 | 0.14 | 0.83 | 0.90 | 0.22 | 26.8 | 4.9 |
| | CE-Net | 0.91 | 0.13 | 0.83 | 0.90 | 0.22 | 21.6 | 4.5 |
| | MultiResUNet | 0.93 | 0.11 | 0.84 | 0.91 | 0.19 | 18.8 | 4.1 |
| | RDAU-NET | 0.91 | 0.11 | 0.84 | 0.91 | 0.20 | 19.3 | 4.1 |
| | SCAN | 0.91 | 0.11 | 0.83 | 0.90 | 0.20 | 26.9 | 4.9 |
| | DenseU-Net | 0.91 | 0.16 | 0.81 | 0.88 | 0.25 | 25.3 | 5.5 |
| | STAN | 0.92 | 0.09 | 0.85 | 0.91 | 0.18 | 18.9 | 3.9 |
| | ESTAN | 0.91 | **0.07** | **0.86** | **0.92** | **0.16** | **16.4** | **3.2** |
| Dataset B [13] | AlexNet | **0.87** | 1.17 | 0.47 | 0.61 | 1.30 | 40.8 | 14.5 |
| | SegNet | 0.85 | 0.83 | 0.60 | 0.71 | 0.98 | 41.6 | 11.4 |
| | U-Net | 0.78 | 0.41 | 0.65 | 0.75 | 0.63 | 39.6 | 10.8 |
| | CE-Net | 0.74 | 0.48 | 0.61 | 0.72 | 0.74 | 40.1 | 10.5 |
| | MultiResUNet | 0.79 | 0.26 | 0.66 | 0.75 | 0.48 | 37.1 | 10.7 |
| | RDAU-NET | 0.78 | 0.30 | 0.67 | 0.77 | 0.52 | 32.4 | 8.3 |
| | SCAN | 0.75 | 0.29 | 0.65 | 0.74 | 0.54 | 43.7 | 9.9 |
| | DenseU-Net | 0.71 | 0.43 | 0.60 | 0.69 | 0.72 | 48.9 | 15.5 |
| | STAN | 0.80 | 0.27 | 0.70 | 0.78 | 0.47 | 35.5 | 9.7 |
| | ESTAN | 0.84 | **0.22** | **0.74** | **0.82** | **0.38** | **25.5** | **7.0** |
| BUSI [29] | AlexNet | **0.87** | 1.14 | 0.55 | 0.68 | 1.27 | 47.4 | 14.1 |
| | SegNet | 0.77 | 0.55 | 0.62 | 0.72 | 0.78 | 46.5 | 13.3 |
| | U-Net | 0.77 | 0.56 | 0.63 | 0.73 | 0.78 | 59.0 | 13.7 |
| | CE-Net | 0.77 | 0.64 | 0.64 | 0.73 | 0.88 | 43.9 | 12.4 |
| | MultiResUNet | 0.78 | 0.37 | 0.67 | 0.75 | 0.59 | 41.2 | 12.0 |
| | RDAU-NET | 0.80 | 0.42 | 0.68 | 0.76 | 0.62 | 39.2 | 12.0 |
| | SCAN | 0.73 | 0.43 | 0.63 | 0.72 | 0.70 | 47.0 | 13.8 |
| | DenseU-Net | 0.74 | 0.43 | 0.64 | 0.72 | 0.69 | 47.4 | 15.5 |
| | STAN | 0.76 | 0.42 | 0.66 | 0.75 | 0.66 | 46.5 | 12.1 |
| | ESTAN | 0.80 | **0.36** | **0.70** | **0.78** | **0.56** | **34.8** | **9.9** |

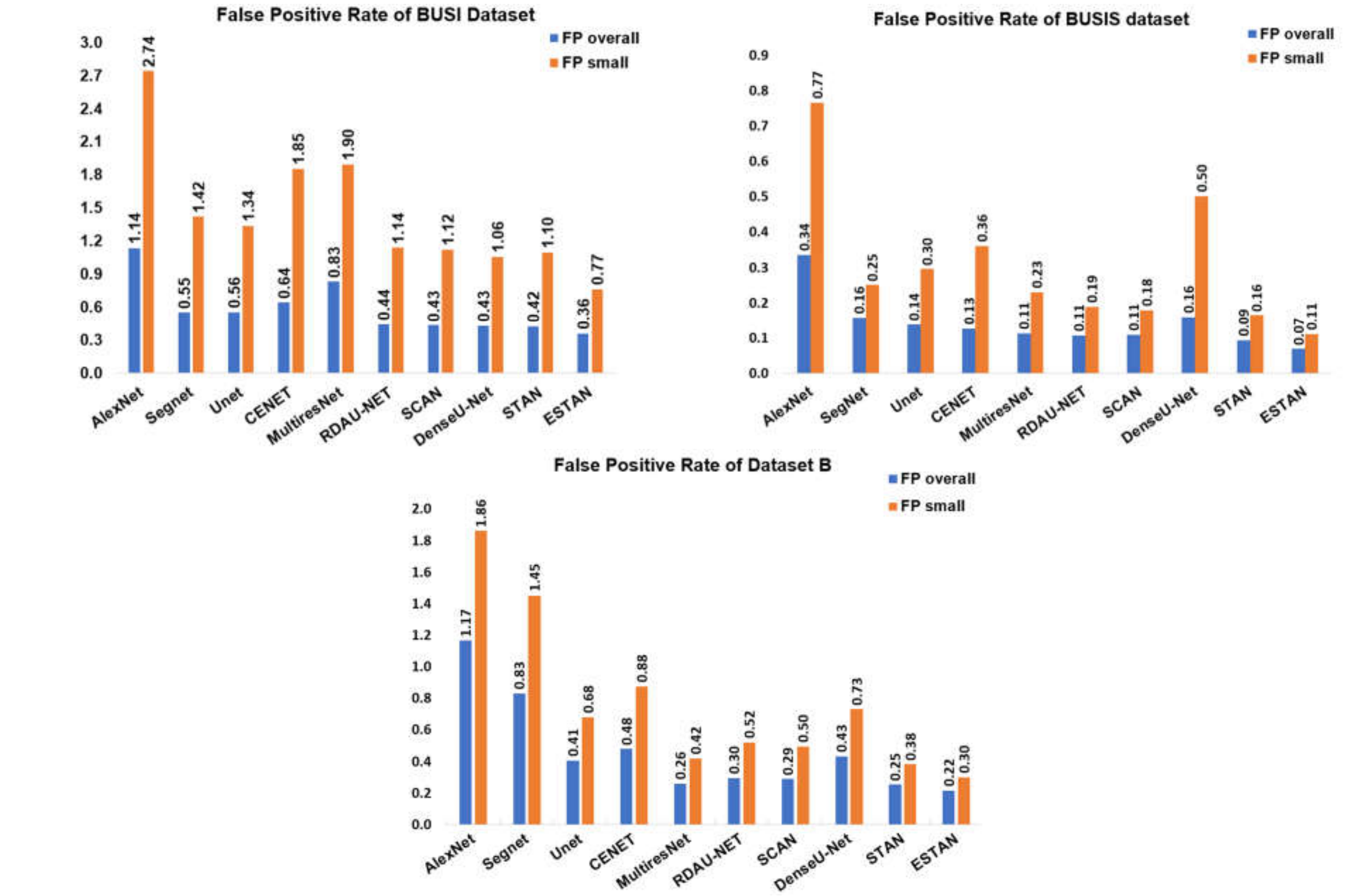

Fig. 6. False positive rates of overall and small tumor segmentation on the three datasets.

### D. Segmenting Tumors with Different Sizes

To demonstrate the effectiveness of the proposed ESTAN model, we split the BUSIS [28] dataset into four tumor size groups. We chose BUSIS dataset for the following reasons: 1) The BUSIS dataset is collected from three hospitals using five ultrasound devices operated by different radiologists; 2) the ground truth of the BUSIS dataset has less bias because it is prepared by four experienced radiologists, where three radiologists generate tumor boundaries for each BUS image separately, and the fourth radiologist—a senior expert—judges and adjusts the majority voting results; and 3) all ten approaches have achieved their best results on BUSIS dataset compared to BUSI and Dataset B. We choose the length of the longest axis of a tumor as our condition to select tumor groups in the original BUS image. The first group contains 19 images with tumor sizes from 0 to 100 pixels, the second group has 30 images from 100 to 120 pixels, the third group consists of 81 images from 120 to 160 pixels, and the fourth group has 432 images from 160 to 533 pixels.

TABLE III
PERFORMANCE OF SMALL TUMOR SEGMENTATION

| Datasets | Methods | TPR | FPR | JI | DSC | AER | AHE | AME |
|---|---|---|---|---|---|---|---|---|
| BUSIS [28] | AlexNet | **0.95** | 0.77 | 0.60 | 0.73 | 0.82 | 26.3 | 9.6 |
| | SegNet | 0.92 | 0.25 | 0.75 | 0.84 | 0.33 | 22.4 | 6.2 |
| | U-Net | 0.92 | 0.30 | 0.76 | 0.84 | 0.38 | 44.2 | 8.3 |
| | CE-Net | 0.91 | 0.36 | 0.73 | 0.82 | 0.46 | 34.8 | 9.0 |
| | MultiResUNet | 0.91 | 0.23 | 0.77 | 0.84 | 0.33 | 27.7 | 8.5 |
| | RDAU-NET | 0.89 | 0.19 | 0.78 | 0.86 | 0.30 | 22.0 | 7.3 |
| | SCAN | 0.88 | 0.18 | 0.77 | 0.85 | 0.30 | 27.4 | 6.2 |
| | DenseU-Net | 0.90 | 0.50 | 0.72 | 0.81 | 0.60 | 34.5 | 8.2 |
| | STAN | 0.90 | 0.17 | 0.79 | 0.87 | 0.26 | 21.3 | 5.2 |
| | ESTAN | 0.90 | **0.11** | **0.82** | **0.89** | **0.21** | **14.9** | **3.0** |
| Dataset B [13] | AlexNet | **0.87** | 1.86 | 0.35 | 0.49 | 2.00 | 49.2 | 18.4 |
| | SegNet | 0.85 | 1.45 | 0.50 | 0.62 | 1.60 | 50.1 | 14.2 |
| | U-Net | 0.77 | 0.68 | 0.59 | 0.68 | 0.91 | 43.1 | 13.8 |
| | CE-Net | 0.72 | 0.88 | 0.53 | 0.63 | 1.15 | 50.0 | 14.4 |
| | MultiResUNet | 0.79 | 0.42 | 0.62 | 0.71 | 0.62 | 39.3 | 11.5 |
| | RDAU-NET | 0.78 | 0.52 | 0.62 | 0.71 | 0.73 | 34.1 | 8.8 |
| | SCAN | 0.75 | 0.50 | 0.61 | 0.70 | 0.74 | 48.7 | 11.2 |
| | DenseU-Net | 0.70 | 0.73 | 0.54 | 0.63 | 1.02 | 56.0 | 20.0 |
| | STAN | 0.81 | 0.40 | 0.67 | 0.76 | 0.59 | 35.9 | 11.1 |
| | ESTAN | 0.85 | **0.30** | **0.72** | **0.80** | **0.44** | **21.5** | **6.3** |
| BUSI [29] | AlexNet | **0.94** | 2.74 | 0.41 | 0.56 | 2.81 | 52.5 | 15.4 |
| | SegNet | 0.81 | 1.42 | 0.55 | 0.66 | 1.61 | 52.1 | 16.6 |
| | U-Net | 0.86 | 1.34 | 0.63 | 0.73 | 1.48 | 61.0 | 13.0 |
| | CE-Net | 0.83 | 1.86 | 0.59 | 0.69 | 2.03 | 50.9 | 13.3 |
| | MultiResUNet | 0.85 | 0.83 | 0.67 | 0.76 | 0.99 | 34.7 | 10.6 |
| | RDAU-NET | 0.87 | 0.99 | 0.68 | 0.77 | 1.13 | 33.9 | 9.9 |
| | SCAN | 0.80 | 1.13 | 0.63 | 0.73 | 1.33 | 42.4 | 12.5 |
| | DenseU-Net | 0.81 | 1.06 | 0.65 | 0.73 | 1.26 | 40.9 | 13.7 |
| | STAN | 0.86 | 1.10 | 0.67 | 0.76 | 1.25 | 49.2 | 11.3 |
| | ESTAN | 0.89 | **0.77** | **0.72** | **0.81** | **0.88** | **24.2** | **6.1** |

TABLE IV
PERFORMANCE OF FOUR TUMOR SIZE GROUPS OF BUSIS DATASET

| Tumor size groups | (0-100) | | (100-120) | | (120-160) | | (>160) | |
|---|---|---|---|---|---|---|---|---|
| Number of Images | 19 | | 30 | | 81 | | 432 | |
| | JI | FP | JI | FP | JI | FP | JI | FP |
| AlexNet | 0.57 | 0.97 | 0.63 | 0.64 | 0.68 | 0.44 | 0.76 | 0.27 |
| SegNet | 0.71 | 0.28 | 0.77 | 0.23 | 0.79 | 0.21 | 0.83 | 0.14 |
| U-Net | 0.72 | 0.34 | 0.78 | 0.27 | 0.80 | 0.18 | 0.84 | 0.11 |
| CE-Net | 0.62 | 0.63 | 0.80 | 0.19 | 0.80 | 0.16 | 0.84 | 0.09 |
| MultiResUNet | 0.71 | 0.34 | 0.80 | 0.16 | 0.82 | 0.17 | 0.86 | 0.09 |
| RDAU-NET | 0.72 | 0.26 | 0.82 | 0.14 | 0.81 | 0.17 | 0.85 | 0.09 |
| SCAN | 0.71 | 0.24 | 0.81 | 0.14 | 0.81 | 0.16 | 0.80 | 0.09 |
| DenseU-Net | 0.67 | 0.77 | 0.75 | 0.34 | 0.78 | 0.21 | 0.83 | 0.11 |
| STAN | 0.76 | 0.25 | 0.81 | 0.11 | 0.83 | 0.12 | 0.86 | 0.08 |
| ESTAN | **0.79** | **0.15** | **0.83** | **0.09** | **0.85** | **0.10** | **0.87** | **0.06** |

TABLE IV lists the results of JIs and FPRs of four tumor groups. AlexNet shows poor performance for segmenting small tumor group with JI of 0.57 and FP of 0.97, while the FP and JI improve dramatically in other three groups. The results of segmenting tumors in both groups (100-120) and (120-160) are very close to each other, e.g., CE-NET and SCAN have achieved the same JI with 0.81 and 0.80 in both groups, respectively. The results show that the tumor size between (0-100) are the most difficult cases, and all ten approaches cannot achieve as good performance as segmenting large tumors. On the other hand, the fourth group contains the large tumor sizes, and all approaches achieve better results than the other tumor size groups. The proposed ESTAN achieves the highest JIs and lowest FPRs on all tumor size groups.

## IV. CONCLUSION

In this paper, we propose the Enhanced Small Tumor-Aware Network (ESTAN) for tumor segmentation in BUS images. ESTAN comprises of two encoder branches that extract and fuse image context information at different scales. The ESTAN blocks apply row-column-wise kernels to adapt to the breast anatomy. The decoder has three skip connections from the two encoders to fuse features. The proposed architecture is sensitive to small breast tumors, and segments small tumor accurately with low false positive rate. In addition, the approach achieves state-of-the-art performance in segmenting tumors with different sizes. We validate the proposed approach extensively using three datasets and compare it with other nine breast tumor segmentation approaches. The results demonstrate that ESTAN achieves the state-of-the-art performance on all datasets.

In the future, we plan to test the proposed approach using large datasets and focus on developing domain-enriched deep architectures for small object detection.

## REFERENCES

[1] JL Ruhl, C. Callaghan, H. A, L. Ries, P Adamo, L. Dickie, and N. Schussler, “Summary Stage 2018: Codes and Coding Instructions,National Cancer Institute,” Bethesda, MD, 2018.

[2] Y. Ikedo, D. Fukuoka, T. Hara, H. Fujita, E. Takada, T. Endo, and T. Morita, “Development of a fully automatic scheme for detection of masses in whole breast ultrasound images,” *Med. Phys.*, vol. 34, no. 11, pp. 4378–4388, 2007.

[3] R. F. Chang, W. J. Wu, W. K. Moon, and D. R. Chen, “Automatic ultrasound segmentation and morphology based diagnosis of solid breast tumors,” *Breast Cancer Res. Treat.*, vol. 89, no. 2, pp. 179–185, 2005.

[4] M. H. Yap, E. A. Edirisinghe, and H. E. Bez, “A novel algorithm for initial lesion detection in ultrasound breast images,” *J. Appl. Clin. Med. Phys.*, vol. 9, no. 4, pp. 181–199, 2008.

[5] S. Joo, Y. S. Yang, W. K. Moon, and H. C. Kim, “Computer-aided diagnosis of solid breast nodules: Use of an artificial neural network based on multiple sonographic features,” *IEEE Trans. Med. Imaging*, vol. 23, no. 10, pp. 1292–1300, 2004.

[6] J. Shan, H. D. Cheng, and Y. Wang, “A novel automatic seed point selection algorithm for breast ultrasound images,” *Proc. - Int. Conf. Pattern Recognit.*, 2008.

[7] M. Xian, Y. Zhang, and H. D. Cheng, “Fully automatic segmentation of breast ultrasound images based on breast characteristics in space and frequency domains,” *Pattern Recognit.*, vol. 48, no. 2, pp. 485–497, 2015.

[8] A. Madabhushi and D. N. Metaxas, “Combining low-, high-level and empirical domain knowledge for automated segmentation of ultrasonic breast lesions,” *IEEE Trans. Med. Imaging*, vol. 22, no. 2, pp. 155–169, 2003.

[9] S. Roy and I. J. Cox, “A Maximum-Flow Formulation of the N-camera Stereo Correspondence Problem,” pp. 492–499.

[10] Y. L. Huang and D. R. Chen, “Automatic contouring for breast tumors in 2-D sonography,” *Annu. Int. Conf. IEEE Eng. Med. Biol. - Proc.*, vol. 7 VOLS, pp. 3225–3228, 2005.

[11] M. Xian, Y. Zhang, H. D. Cheng, F. Xu, B. Zhang, and J. Ding, “Automatic breast ultrasound image segmentation: A survey,” *Pattern Recognit.*, vol. 79, pp. 340–355, 2018.

[12] K. Huang, Y. Zhang, H. D. Cheng, P. Xing, and B. Zhang, “Fuzzy Semantic Segmentation of Breast Ultrasound Image with Breast Anatomy Constraints,” pp. 1–15, 2019.

[13] M. H. Yap, G. Pons, J. Martí, S. Ganau, M. Sentís, R. Zwiggelaar, A. K. Davison, and R. Martí, “Automated Breast Ultrasound Lesions Detection Using Convolutional Neural Networks,” *IEEE J. Biomed. Heal. Informatics*, vol. 22, no. 4, pp. 1218–1226, 2018.

[14] Z. Zhuang, N. Li, A. N. J. Raj, V. G. V. Mahesh, and S. Qiu, “An RDAU-NET model for lesion segmentation in breast ultrasound images,” *PLoS One*, vol. 14, no. 8, pp. 1–23, 2019.

[15] Y. Hu, Y. Guo, Y. Wang, J. Yu, J. Li, S. Zhou, and C. Chang, “Automatic tumor segmentation in breast ultrasound images using a dilated fully convolutional network combined with an active contour model,” *Med. Phys.*, vol. 46, no. 1, pp. 215–228, 2019.

[16] A. Vakanski, M. Xian, and P. E. Freer, “Attention Enriched Deep Learning Model for Breast Tumor Segmentation in Ultrasound Images,” 2019.

[17] M. Byra, P. Jarosik, K. Dobruch-Sobczak, Z. Klimonda, H. Piotrzkowska-Wroblewska, J. Litniewski, and A. Nowicki, “Breast mass segmentation based on ultrasonic entropy maps and attention gated U-Net,” 2020.

[18] W. K. Moon, Y. W. Lee, H. H. Ke, S. H. Lee, C. S. Huang,

and R. F. Chang, “Computer-aided diagnosis of breast ultrasound images using ensemble learning from convolutional neural networks,” *Comput. Methods Programs Biomed.*, vol. 190, 2020.
[19] H. Lee, J. Park, and J. Y. Hwang, “Channel Attention Module with Multi-scale Grid Average Pooling for Breast Cancer Segmentation in an Ultrasound Image,” *IEEE Trans. Ultrason. Ferroelectr. Freq. Control*, vol. 3010, no. c, pp. 1–1, 2020.
[20] J. Chen, Chenyi; Liu, Ming-Yu; Tuzel, C. Oncel; Xiao, “R-CNN for Small Object Detection,” *Asian Conf. Comput. Vis.*, pp. 214–230, 2016.
[21] H. Krishna and C. V. Jawahar, “Improving small object detection,” *Proc. - 4th Asian Conf. Pattern Recognition, ACPR 2017*, pp. 346–351, 2018.
[22] L. Guan, Y. Wu, and J. Zhao, “SCAN: Semantic context aware network for accurate small object detection,” *Int. J. Comput. Intell. Syst.*, vol. 11, no. 1, pp. 951–961, May 2018.
[23] R. Dong, X. Pan, and F. Li, “DenseU-Net-Based Semantic Segmentation of Small Objects in Urban Remote Sensing Images,” *IEEE Access*, vol. 7, pp. 65347–65356, 2019.
[24] B. Shareef, M. Xian, and A. Vakanski, “STAN : Small Tumor-Aware Network for Breast Ultrasound Image Segmentation,” *IEEE Int. Symp. Biomed. Imaging (ISBI 2020)*, 2020.
[25] J. Araujo, André and Norris, Wade and Sim, “Computing Receptive Fields of Convolutional Neural Networks,” *Distill*, 2019.
[26] W. Luo, Y. Li, R. Urtasun, and R. Zemel, “Understanding the effective receptive field in deep convolutional neural networks,” *Adv. Neural Inf. Process. Syst.*, no. Nips, pp. 4905–4913, 2016.
[27] F. Yu and V. Koltun, “Multi-scale context aggregation by dilated convolutions,” *4th Int. Conf. Learn. Represent. ICLR 2016 - Conf. Track Proc.*, 2016.
[28] M. Xian, Y. Zhang, H. D. Cheng, F. Xu, K. Huang, B. Zhang, J. Ding, C. Ning, and Y. Wang, “A Benchmark for Breast Ultrasound Image Segmentation (BUSIS),” pp. 1–9, 2018.
[29] W. Al-Dhabyani, M. Gomaa, H. Khaled, and A. Fahmy, “Dataset of breast ultrasound images,” *Data Br.*, vol. 28, p. 104863, 2020.
[30] D. P. Kingma and J. Ba, “Adam: A Method for Stochastic Optimization,” pp. 1–15, 2014.
[31] F. Milletari, N. Navab, and S. A. Ahmadi, “V-Net: Fully convolutional neural networks for volumetric medical image segmentation,” *Proc. - 2016 4th Int. Conf. 3D Vision, 3DV 2016*, pp. 565–571, 2016.
[32] E. Shelhamer, J. Long, and T. Darrell, “Fully Convolutional Networks for Semantic Segmentation,” 2017.
[33] V. Badrinarayanan, A. Kendall, and R. Cipolla, “SegNet: A Deep Convolutional Encoder-Decoder Architecture for Image Segmentation,” *IEEE Trans. Pattern Anal. Mach. Intell.*, vol. 39, no. 12, pp. 2481–2495, Dec. 2017.
[34] O. Ronneberger, P. Fischer, and T. Brox, “U-net: Convolutional networks for biomedical image segmentation,” *Int. Conf. Med. image Comput. Comput. Interv. O., Fischer, P., Brox, T. (2015, October). Springer, Cham.*, vol. 9351, pp. 234–241, 2015.
[35] Z. Gu, J. Cheng, H. Fu, K. Zhou, H. Hao, Y. Zhao, T. Zhang, S. Gao, and J. Liu, “CE-Net: Context Encoder Network for 2D Medical Image Segmentation,” *IEEE Trans. Med. Imaging*, vol. 38, no. 10, pp. 2281–2292, 2019.
[36] N. Ibtehaz and M. S. Rahman, “MultiResUNet : Rethinking the U-Net architecture for multimodal biomedical image segmentation,” *Neural Networks*, vol. 121, pp. 74–87, 2020.
[37] R. Dong, X. Pan, and F. Li, “DenseU-Net-Based Semantic Segmentation of Small Objects in Urban Remote Sensing Images,” *IEEE Access*, vol. 7, pp. 65347–65356, 2019.